# OIG and SARG CCDs Characterization

S. SCUDERI[1], G. BONANNO[1], P. BRUNO[1], A. CALI[1], R. COSENTINO[2]

[1]Osservatorio Astrofisico di Catania, Catania, Italy

[2]Telescopio Nazionale Galileo, Canary Islands, Spain

## 1. Introduction

CCDs characterization is the preliminary step to perform before the CCD can be properly used at the telescope.

Most of the scientific instrumentation at the Italian National Telescope "Galileo" use CCDs as detectors. In particular the optical imager (OIG) and the high resolution spectrograph (SARG) use a mosaic of two 2k x 4k CCD manufactured by EEV (EEV 4280). The technical characteristics of the EEV4280 can be found in Cosentino et al (these proceedings).

## 2. The Calibration Facility

At the Catania Astrophysical Observatory we have developed a facility that allows the full electro-optical characterization of bidimensional detectors.

The main component of the facility is the instrumental apparatus to measure the quantum efficiency of the detector in the wavelength range 1250 - 11000 Å. The system has a modular structure. The first module accomodates the lamps (xenon and deuterium) and an optical system (made up of mirrors and diaphragms) to match the focal ratio (f/5.4) of the monochromator. The second module is the monochromator. It has a Czerny-Turner configuration and is equipped with three 1200 g/mm ruled gratings. The third module is a camera containing a $MgF_2$ beam splitter in which the radiation beam, after being dispersed, is divided into a reflected and transmitted beam. The last module is made by two twin cameras which, through a $MgF_2$ lens, focus the beam on the CCD and on the calibrated detector. Under 2000 Å the system is operated under vacuum conditions. A more detailed description of the system can be found in Bonanno et al 1996.

For uniformity and linearity measurements a 20 inches integrating sphere is optically connected to the QE measurement system through a quartz singlet. The useful wavelength interval range in this case is 2000 - 11000 Å.

The system gain measurements and the CTE analysis are performed using a camera accomodating a $Fe^{55}$ x-ray source which is connected to the CCD dewar.

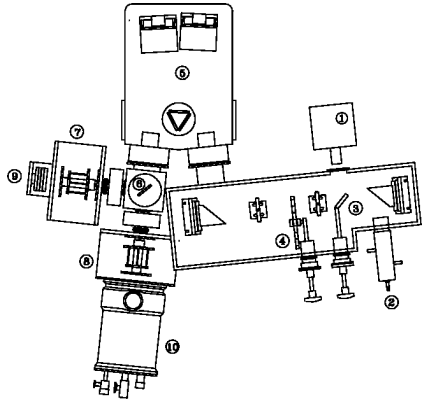

Fig. 1. The instrumental apparatus for the measurement of the quantum efficiency: 1) xenon lamp, 2) deuterium lamp, 3) moving plane mirror, 4) filter wheel, 5) monochromator, 6) beam splitter camera, 7) photodiode camera, 8) CCD camera, 9) photodiode, 10) CCD cryostat.

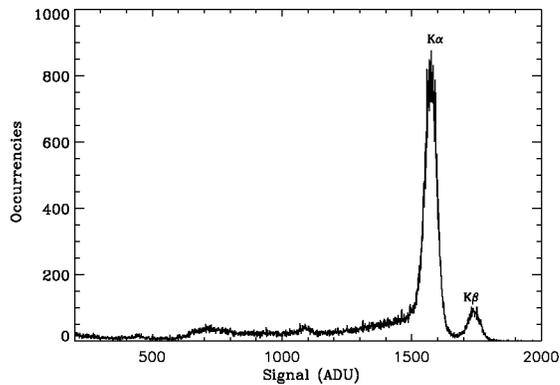

Fig. 2. X-ray spectrum of a radioactive $Fe^{55}$ source. The K$\alpha$ (5.9KeV) and the K$\beta$ (6.5keV) lines are clearly visible. Only single pixel events are shown in the figure.

## 3. The measurements

3.1. System gain measurement

The measurement of the system gain is obtained exposing the CCD to a $Fe^{55}$ X-ray source. The main line (K$\alpha$) emitted by the source has an energy of 5.9 keV, so the charge deposited by a single photon in each pixel is 1620 $e^-$. The measurement of the position of the line on the spectrum of the source recorded by the CCD allows to determine directly the system gain with a precision better than 1%. An example of the X-ray spectrum of the $Fe^{55}$ source is shown in figure 2 for the OIG CCD.

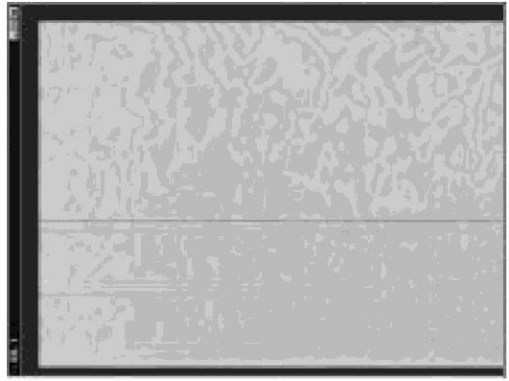

Fig. 3. Flat field of a OIG CCD obtained at 9000 Å. Interference fringes are clearly visible.

## 3.2. Linearity Analysis

The linearity is measured illuminating the CCD with a uniform source of radiation at increasing exposure times. At each exposure time, $t_i$, we compute the average signal, $S_i$, in a given area and define the deviation from linearity at that signal level as:

$$Deviation\ from\ linearity = \frac{S_i/t_i}{Sm} - 1$$

where Sm is the average signal per time unit defines as $Sm = \frac{1}{n}\sum_i \frac{S_i}{t_i} - 1$ where the sum is over the n exposures.

## 3.3. Dark Current Analysis

The dark current is obtained through the following procedure:

$$Difference = Dark_{long} - Dark_{short}$$

where $Dark_{Long}$ and $Dark_{short}$ are 1 hour and 5 min dark exposures respectively. The frame Difference is divided in 10x10 pixels sub frames ($box_i$) for which the dark current is computed individually. The value of the dark current is obtained as the average of these values.

$$Dark\ Current = mean(box_i(Difference))$$

## 3.4. Uniformity analisys

The uniformity of the CCD is measured by illuminating the detector with a uniform source of radiation at different wavelengths. The deviation from homogeneity in a given area is obtained as:

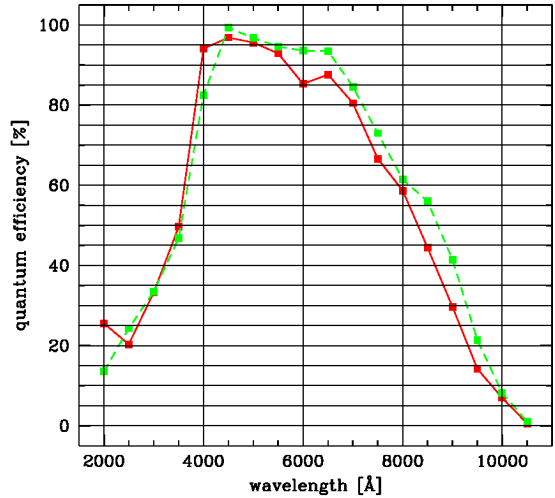

Fig. 4. The quantum efficiency of the OIG CCD (grey dashed line) and of the SARG CCD (black solid line)

$$\text{Deviation from Homogeneity} = \frac{RMS(area)}{Mean(area)} \ [\%]$$

Values of deviation from homogeneity are given in table 1 at three different wavelengths for the two CCDs.

### 3.5. Quantum Efficiency

The quantum efficiency is measured illuminating the CCD with monochromatic radiation (see the panel that describes the QE calibration facility) and comparing its response to that of a calibrated detector. The measurements were done between 2000 and 10500 Å in incremental step of 500 Å. In Fig. 4 we have plotted the quantum efficiency of the OIG CCD and of the SARG CCD.

### 3.6. Charge Transfer Efficiency

The charge transfer efficiency (CTE) is measured using the X-ray stimulation method. In particular, the parallel CTE is obtained exposing the CCD to a $Fe^{55}$ source for a given amount of time. After integration, the columns are stacked together and the signal is plotted versus the number of pixel transfers. The CTE is then given by:

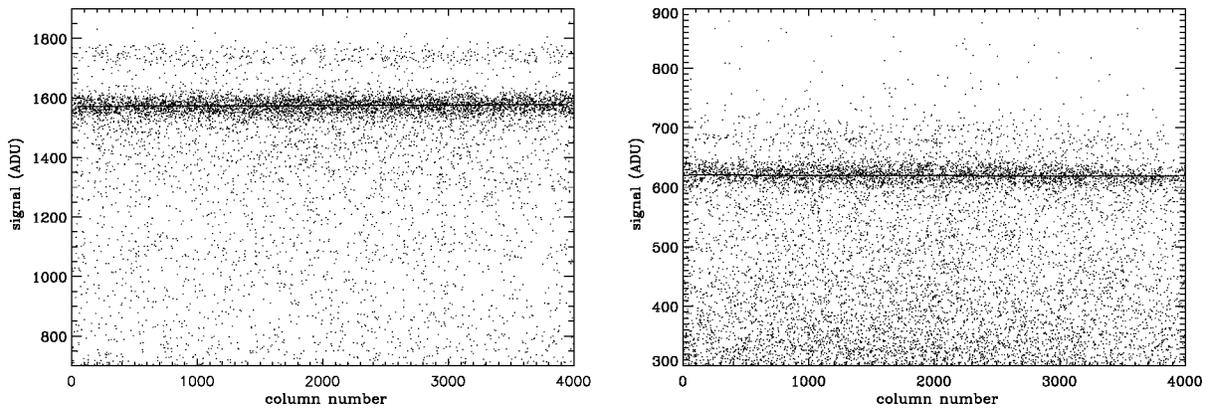

Fig. 5. The Fe$^{55}$ X-ray response in the vertical direction for the OIG CCD (left) and the SARG CCD (right).

$$CTE = 1 - \frac{Charge\ Loss}{(1620\ e-)(Nt)}$$

where charge loss is the amount of charge lost in Nt transfers.

Figure 5 shows the results of this analysis for the OIG CCD (Fig. 5a) and for the SARG CCD. In both cases the CTE is very good (see table 1) and within the specifications of the manufacturer.

Tab. 1 - CCD Characteristics: Summary

|  |  | EEV4280 - SARG | EEV 4280 - OIG |
|---|---|---|---|
| Linearity |  | <3% | <0.5% |
| Dark Current |  | 7 e pix$^{-1}$h$^{-1}$ | 6 e pix$^{-1}$h$^{-1}$ |
| Uniformity at: | 4000 Å | 1.5% | 3.9% |
|  | 6000 Å | 2.2% | 3.4% |
|  | 9000 Å | 2.6% | 5.2% |
| QE at: | 4000 Å | 94% | 82% |
|  | 6000 Å | 85% | 94% |
|  | 9000 Å | 30% | 41% |
| CTE |  | 0.999998 | 0.999999 |